\documentclass[aps,prb,twocolumn,superscriptaddress]{revtex4-1}

\usepackage{amssymb}
\usepackage{amsmath}
\usepackage{color}
\usepackage{bm}
\usepackage{epsfig}
\usepackage{graphicx}
\usepackage{dcolumn}
\usepackage[pagewise,columnwise]{lineno}
\usepackage{setspace}
\usepackage[utf8]{inputenc}
\usepackage{color}
\usepackage{nicefrac}
\usepackage{hyperref}
\usepackage{color}

\begin{document}


\title{Magnetic and electronic properties at the $\gamma$-Al$_2$O$_3$/SrTiO$_3$ interface}


\author{J. R. L. Mardegan}
\email[]{jrlmardegan@gmail.com}
\affiliation{Swiss Light Source, Paul Scherrer Institut, 5232 Villigen PSI, Switzerland}

\author{D. V. Christensen}
\email[]{dechr@dtu.dk}
\affiliation{Technical University of Denmark, Department of Energy Conversion and Storage, Risø campus, Roskilde, Denmark}

\author{Y. Z. Chen}
\affiliation{Technical University of Denmark, Department of Energy Conversion and Storage, Risø campus, Roskilde, Denmark}

\author{S. Parchenko}
\affiliation{Swiss Light Source, Paul Scherrer Institut, 5232 Villigen PSI, Switzerland}

\author{S. R. V. Avula}
\affiliation{Swiss Light Source, Paul Scherrer Institut, 5232 Villigen PSI, Switzerland}

\author{N. Ortiz-Hernandez}
\affiliation{Swiss Light Source, Paul Scherrer Institut, 5232 Villigen PSI, Switzerland}

\author{M. Decker}
\affiliation{Swiss Light Source, Paul Scherrer Institut, 5232 Villigen PSI, Switzerland}

\author{C. Piamonteze}
\affiliation{Swiss Light Source, Paul Scherrer Institut, 5232 Villigen PSI, Switzerland}

\author{N. Pryds}
\affiliation{Technical University of Denmark, Department of Energy Conversion and Storage, Risø campus, Roskilde, Denmark}

\author{U. Staub}
\email[]{urs.staub@psi.ch}
\affiliation{Swiss Light Source, Paul Scherrer Institut, 5232 Villigen PSI, Switzerland}

\date{\today}

\begin{abstract}

The magnetic and electronic nature of the $\gamma$-Al$_2$O$_3$/SrTiO$_3$ spinel/perovskite interface is explored by means of x-ray absorption spectroscopy.  
Polarized x-ray techniques combined with atomic multiplet calculations reveal localized magnetic moments assigned to Ti$^{3+}$ at the interface with equivalent size for in- and out-of-plane magnetic field directions.    
Although magnetic fingerprints are revealed, the Ti$^{3+}$ magnetism can be explained by a paramagnetic response at low temperature under applied magnetic fields.  
Modeling the x-ray linear dichroism results in a $\Delta_0 \sim$ 1.9 eV splitting between the $t_{2g}$ and $e_g$ states for the Ti$^{4+}$ 3$d^0$ orbitals. In addition these results indicate that the lowest energy states have the out-of-plane $d_{xz}/d_{yz}$ symmetry.       
The isotropic magnetic moment behavior and the lowest energy $d_{xz}/d_{yz}$ states are in contrast to the observations for the two-dimensional electron gas at the perovskite/perovskite interface of LaAlO$_3$/SrTiO$_3$, that exhibits an anisotropic magnetic $d_{xy}$ ground state. 

\end{abstract}

\pacs{71.20.Lp, 62.50.-p, 72.20.Pa, 74.62.Fj } 


\maketitle


A complete understanding of the electronic and magnetic properties confined at the interface between two oxide insulators remains a great challenge. Progress in this field is crucial for engineering materials with novel properties.  
Oxide interfaces have recently drawn attention due to an enormous amount of observed physical phenomena, such as the emergence of a two-dimensional electron gas (2DEG),\cite{Hwang_Nature_2004, Thiel_Science_2006,Chen_Nat_Comm_2013} superconductivity,\cite{Reyren_Science_2007, Caviglia_Nature_2008, Dikin_PRL_2011,Ariando_NatComm_2011} magnetism,\cite{Brinkman_NatMat_2007, Lee_NatMat_2013, Salluzzo_PRL_2013} coexistence of magnetism and superconductivity,\cite{Ariando_NatComm_2011, Li_NatPhys_2011, Bert_NatPhys_2011, Gan_AdvMat_2019} Rashba spin-orbit coupling,\cite{Caviglia_PRL_2010} and nematicty.\cite{Davis_PRB_2018}  
In this context, the interface of LaAlO$_3$/SrTiO$_3$ (LAO/STO), created by two  non-magnetic and band insulating perovskites, has been extensively investigated.  
Even with the most recent theoretical and experimental efforts, its interface properties, such as high electron mobility ($\mu_\text{LAO} \sim$ 10$^{3}$ cm$^{2}$/Vs)\cite{Trier_JPDAP_2018} and diverse types of magnetic ordering\cite{Brinkman_NatMat_2007, Barriocanal_NatCommun_2010, Kalisky_NatComm_2012, Salluzzo_PRL_2013, Banerjee_NatMat_2013, Pai_arxiv_2016} remain strongly debated due to controversial findings.  

Similar to the LAO/STO material, a new class of heterostructures composed of two non-magnetic band insulators have shown analogous properties, such as high-electron mobility ($\mu_\text{GAO} \sim$ 1.4 x 10$^{5}$ cm$^{2}$/Vs) \cite{Chen_Nat_Comm_2013, Schutz_PRB_2017, Christensen_PRA_2018} and magnetic fingerprints\cite{Christensen_NatMat_2018} at the interface: $\gamma$-Al$_2$O$_3$/SrTiO$_3$ (GAO/STO).    
Although these interfaces present several similar or sometimes even superior properties compared with LAO/STO,\cite{Christensen_APL_2016, Niu_NanoLett_2017} a full comprehension of the interface properties for the GAO/STO remains lacking.   
In addition, it is surprising that these heterostructures grow epitaxially with materials that exhibits distinct symmetries (GAO/STO: spinel/perovskite).  
Motivated by the unconventional electronic and magnetic properties that are likely to be due to the Ti $t_{2g}$/$e_g$ orbitals at the interface, we carried out a microscopic investigation on this heterostructure.  

X-ray spectroscopy performed on the LAO/STO system\cite{Salluzzo_PRL_2009, Sing_PRL_2009, Zhou_PRB_2011, Salluzzo_AdvMater_2013, Lee_NatMat_2013, Pesquera_PRL_2014} revealed that within the first few nanometers at the STO side of the interface, the Ti ions can adopt two different states, $i.e.$, Ti$^{3+}$ and Ti$^{4+}$ oxidation states. 
This has direct influence on the magnetic properties since the presence of 3$d^1$ (Ti$^{3+}$) bands carry magnetic moments.  
However, for the LAO/STO interface the magnetic response is still under discussion since it was suggested that it has in-plane ferromagnetism\cite{Lee_NatMat_2013} or a complex magnetic structure.\cite{Salluzzo_PRL_2009, Triscone_NatPhys_2013}  
Such magnetic phenomena are also expected to occur in other heterostructures, such as the GAO/STO investigated here.   
Recently, Christensen \textit{et al.}  [\onlinecite{Christensen_NatMat_2018}] performed investigations with scanning superconducting quantum interference device (SQUID) measurements and found magnetic stripes different from the magnetic patches found earlier in LAO/STO.  
In this study, a long-range magnetic order was observed below 40 K as well as a strong tunability of the magnetic signal upon application of stress to the sample. 
However, the scanning SQUID technique only provides local measurements of the magnetic field arising from the magnetic state.  
Therefore, the information on the origin of the magnetism or on ions responsible for it is still missing. 

The electronic configuration of the Ti ions in the LAO/STO heterostructures was found to be formed by in-plane Ti$^{4+}$ 3$d_{xy}$ states capable of hosting the 2DEG and reside in proximity to the interface.  
By contrast, $d_{xz}$ or $d_{yz}$ typically account for the high mobility charges in the system, and reside away from the interface into the STO.\cite{Joshua_NatCommun_2012, Salluzzo_AdvMater_2013} 
Concerning the GAO/STO heterostructures, Cao \textit{et al.} [\onlinecite{Cao_NPJ_2016}] have recently observed that the lowest-lying orbitals are the out-of-plane $d_{xz}/d_{yz}$ orbitals with a strain-tunable splitting between the subbands. This inversion of the Ti 3$d$ subbands are expected to influence the electronic and magnetic properties. 

In this letter, properties of the $\gamma$-Al$_2$O$_3$/SrTiO$_3$ interface were investigated by means of x-ray absorption spectroscopy (XAS) techniques at the Ti $L_{2,3}$ edges.  
 
X-ray linear dichroism (XLD) measurements suggest an orbital configuration in which the out-of-plane Ti $d_{xz}$/$d_{yz}$ subbands are lowest in energy, however, being very close to the in-plane $d_{xy}$ states.
Using x-ray magnetic circular dichroism (XMCD) we observe that the magnetism is mostly due to Ti$^{3+}$ ions present at the interface with a paramagnetic-like response at low temperatures.  

\begin{figure}[!t]
\centering
\includegraphics[trim=0.0cm 0.0cm 0.0cm 0.0cm, clip=true, totalheight=0.25 \textheight, angle=0]{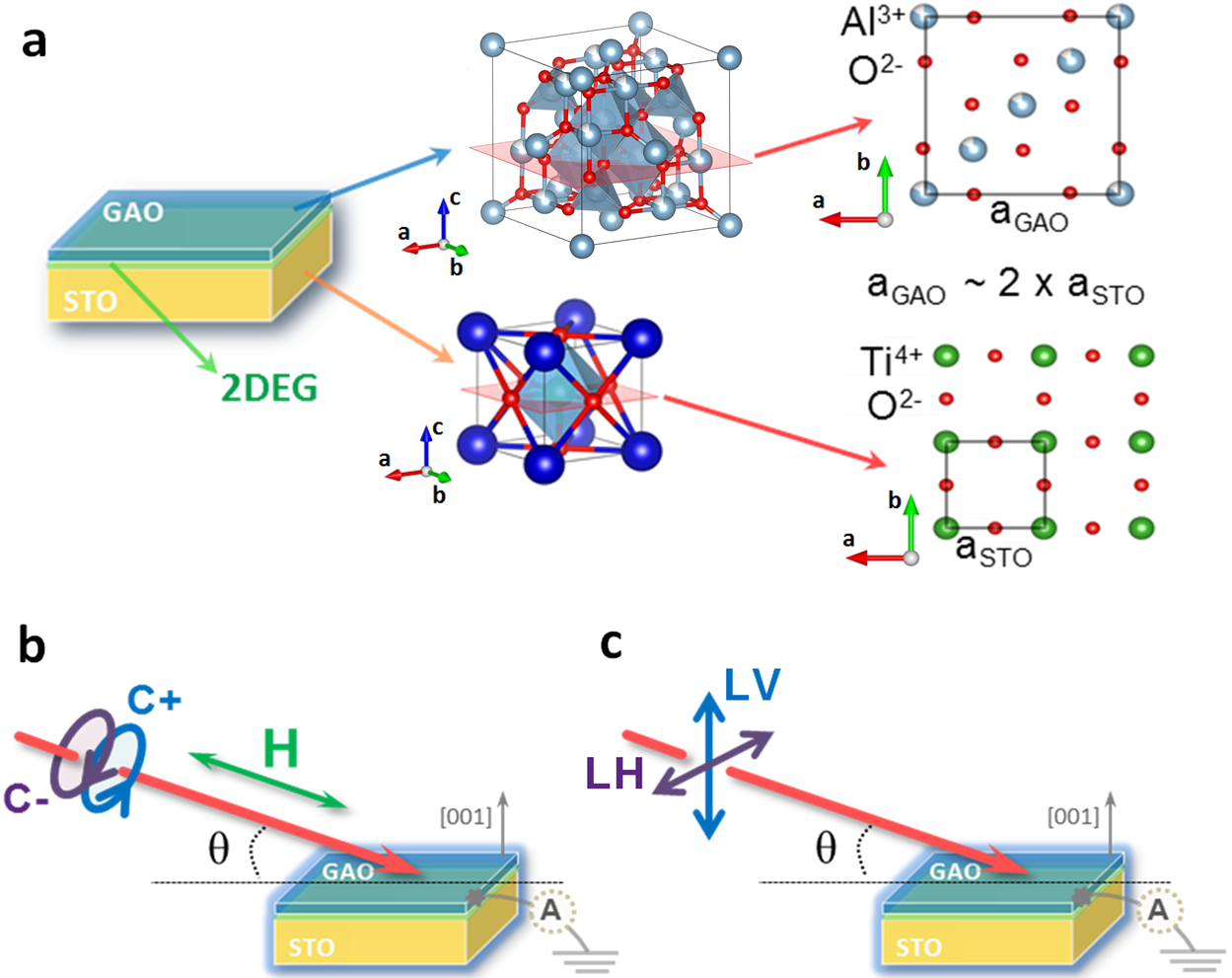}

\caption{(Color online) (a) Schematics of the GAO/STO structure and the formation of the 2DEG at the interface. GAO lattice parameter (a$_{\text{GAO}}$) is approximately twice the STO unit cell (a$_{\text{STO}}$). Illustration of the experimental setup for XMCD and XLD measurements are shown in panels (b) and (c), respectively. An external magnetic field ($\textbf{H}$) is applied parallel or antiparallel to the x-ray wave-vector; XMCD intensity is the difference between absorption spectra obtained with left- ($\mu_{+}$; C-) and right-handed ($\mu_{+}$; C+) circular polarization; XLD intensity is the difference between absorption spectra obtained with linear horizontally ($\mu_{ab}$; LH) and vertically ($\mu_{c}$; LV) polarization. 
}
\label{fig:fig1_setup}
\end{figure}

Spinel $\gamma$-Al$_2$O$_3$ thin films of approximately 3.5 unit cells (u.c.)  $\sim$ 2.7 nm thickness were deposited by pulsed-laser deposition on a perovskite TiO$_2$-terminated (001)-oriented SrTiO$_3$ as discussed elsewhere.\cite{Christensen_PRA_2018}  
Although both materials exhibit different symmetries, the GAO lattice parameter (a$_{\text{GAO}} \sim 7.911$ \AA{}) is almost twice as large as the STO lattice (a$_{\text{STO}} \sim 3.905$ \AA{}),\cite{Gutierrez_PRB_2002, Jiang_JPC_2010} which yielded a compressive strain on the GAO film smaller than 1.3~$\%$.  
A structural representation of the spinel/perovskite material is sketched in Figure \ref{fig:fig1_setup}(a).  
Transport measurements were performed to characterize this heterostructure and current results are in good agreement with recent works.\cite{Christensen_NatMat_2018}  

XAS, XMCD and XLD measurements with the x-rays energy tuned to the Ti $L_{2,3}$ edges (454 - 460 eV) were performed at the XTreme beamline\cite{Piamonteze_JSR_2012} of the Swiss Light Source using the total-electron-yield (TEY) detection mode.    
The experimental geometries used for the absorption measurements with circular and linear polarization are represented in Figures \ref{fig:fig1_setup}(b) and \ref{fig:fig1_setup}(c), respectively.

The magnetic properties of the interface at low temperature ($\sim$ 3 K) and under magnetic field up to 6.5 T were investigated by means of the XMCD technique in two sample configurations\cite{geometry} ($i.e.$, at $\theta = 30^{\circ}$ and $90^{\circ}$ incidence angles as shown in panel~\ref{fig:fig1_setup}(b)).    
X-ray absorption spectra with left- and right-handed circular polarization were subtracted ($\mu_{+}$ - $\mu_{-}$) to provide the XMCD intensity. 
To improve the signal-to-noise ratio and remove non-magnetic contributions, at least 24 absorption spectra with magnetic fields parallel and antiparallel to the incident x-ray wave-vector were collected.  
  
The Ti 3$d$ orbital configurations were investigated using the XLD technique at room temperature and at the two different sample configurations ($\theta = 30^{\circ}$ and $90^{\circ}$ incidence angles: panel~\ref{fig:fig1_setup}(c)).   
Absorption spectra with linearly horizontal and vertical polarized light were recorded, for which the XLD signal is obtained from the difference between the two spectra ($\mu_{c}$ - $\mu_{ab}$).



\begin{figure}[!t]
\centering
\includegraphics[trim=0.0cm 0.0cm 0.0cm 0.0cm, clip=true, totalheight=0.22 \textheight, angle=0]{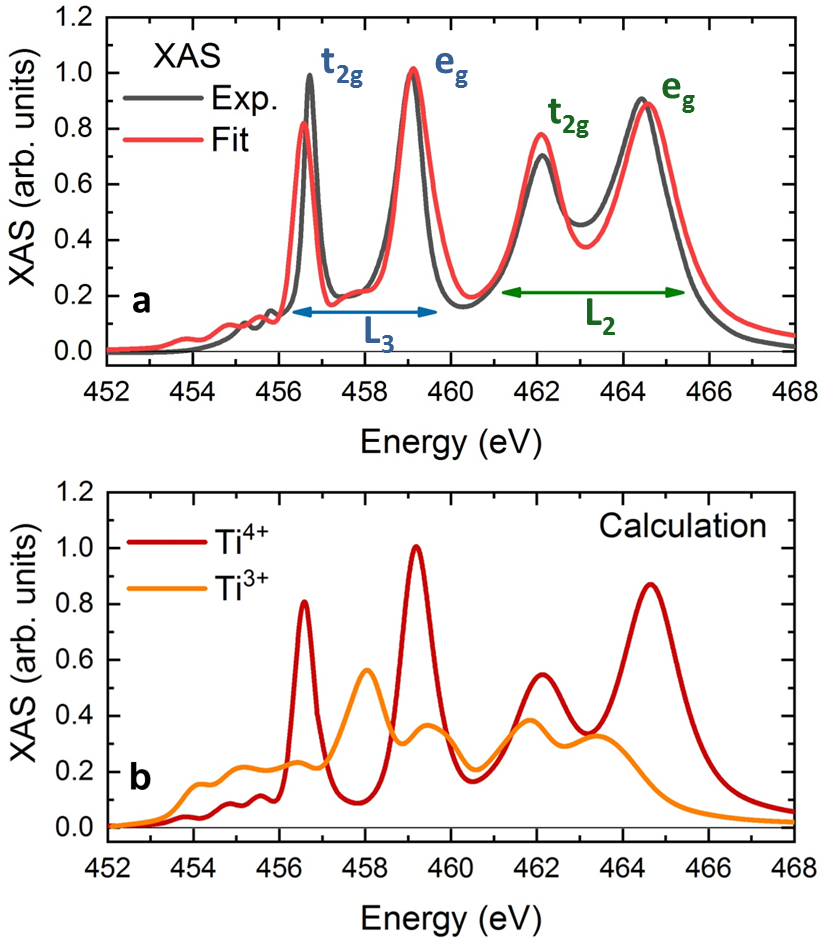}
\caption{(Color online) (a) X-ray absorption spectra collected at the Ti $L_{2,3}$ edges and compared to the linear fits of the Ti$^{4+}$ and Ti$^{3+}$ contributions. The octahedral splitting for the Ti$^{4+}$ 3$d$ bands results in two peaks labeled as $t_{2g}$ and $e_g$ for each edge. The experimental data is obtained averaging two absorption spectra with horizontally and vertically polarized light. Panel (b) shows the calculated spectra for octahedral coordinated Ti$^{4+}$/Ti$^{3+}$ ions which were used in panel (a) to fit the experimental data and to extract their contributions. }
\label{fig:fig2_XAS}
\end{figure}

Figure \ref{fig:fig2_XAS}(a) displays the x-ray absorption spectrum obtained in the vicinity of the Ti $L_{2,3}$ absorption edges for the GAO/STO heterostructure.     
The experimental data displayed in panel \ref{fig:fig2_XAS}(a) was collected averaging two absorption spectra with horizontally and vertically polarized light.   
The clear splitting of the crystal field ($\Delta_0$ in Fig. \ref{fig:fig4_XLD}(d)) into $e_g$ and $t_{2g}$ states is observed at the $L_2$ and $L_3$ edges which represents the Ti$^{4+}$ (3$d^0$) electronic configuration.    
The splitting within the $e_g$ and $t_{2g}$ manifolds ($\Delta_{eg}$ and $\Delta_{t2g}$ in Fig. \ref{fig:fig4_XLD}(d)) into subbands cannot be directly unveiled by the XAS spectra due to the core hole broadening.  
Therefore, we compare the experimental results to atomic multiplet calculations carried out with the \textsc{ctm4xas} code.\cite{Groot_PRB_1990, Stavitski_M_2010}  
These calculations are displayed in Fig.~\ref{fig:fig2_XAS}(b) for octahedrally coordinated Ti$^{4+}$ and Ti$^{3+}$ ions.\cite{code_XAS}     
A fit with a linear combination of the two calculated Ti ions spectra (Fig.~\ref{fig:fig2_XAS}(b)) results in a good agreement with the experimental data (Fig.~\ref{fig:fig2_XAS}(a)). 
The best agreement between calculated and experimental data results in an approximately 10 $\%$ Ti$^{3+}$ ion contribution. 
As it is believed that the 3$d^1$ bands exist only in the first few nm of the interface,\cite{Mele_book_2015} the amount of Ti$^{3+}$ signal will strongly depend on the electron escape depth.   
The TEY method employed here has a probe depth between 1.5 - 3 nm, which makes it mainly sensitive to the GAO/STO interface.\cite{Iyasu_ASS_2006}    
Therefore, assuming that Ti$^{3+}$ ions will be located directly at the interface, approximately 60 \% of the interfacial Ti will be trivalent. 

The occurrence of such 3$d^1$ bands has been argued to be related to a rich amount of oxygen vacancies in these heterostructures.\cite{Christensen_AEM_2017, Gunkel_AAMI_2017}
XAS spectra reported for some LAO/STO heterostructures may indicate a fraction larger than 10 $\%$ of Ti$^{3+}$ at the interface.\cite{Mele_book_2015}

\begin{figure}[!b]
\centering

\includegraphics[trim=0.0cm 0.0cm 0.0cm 0.0cm, clip=true, totalheight=0.215 \textheight, angle=0]{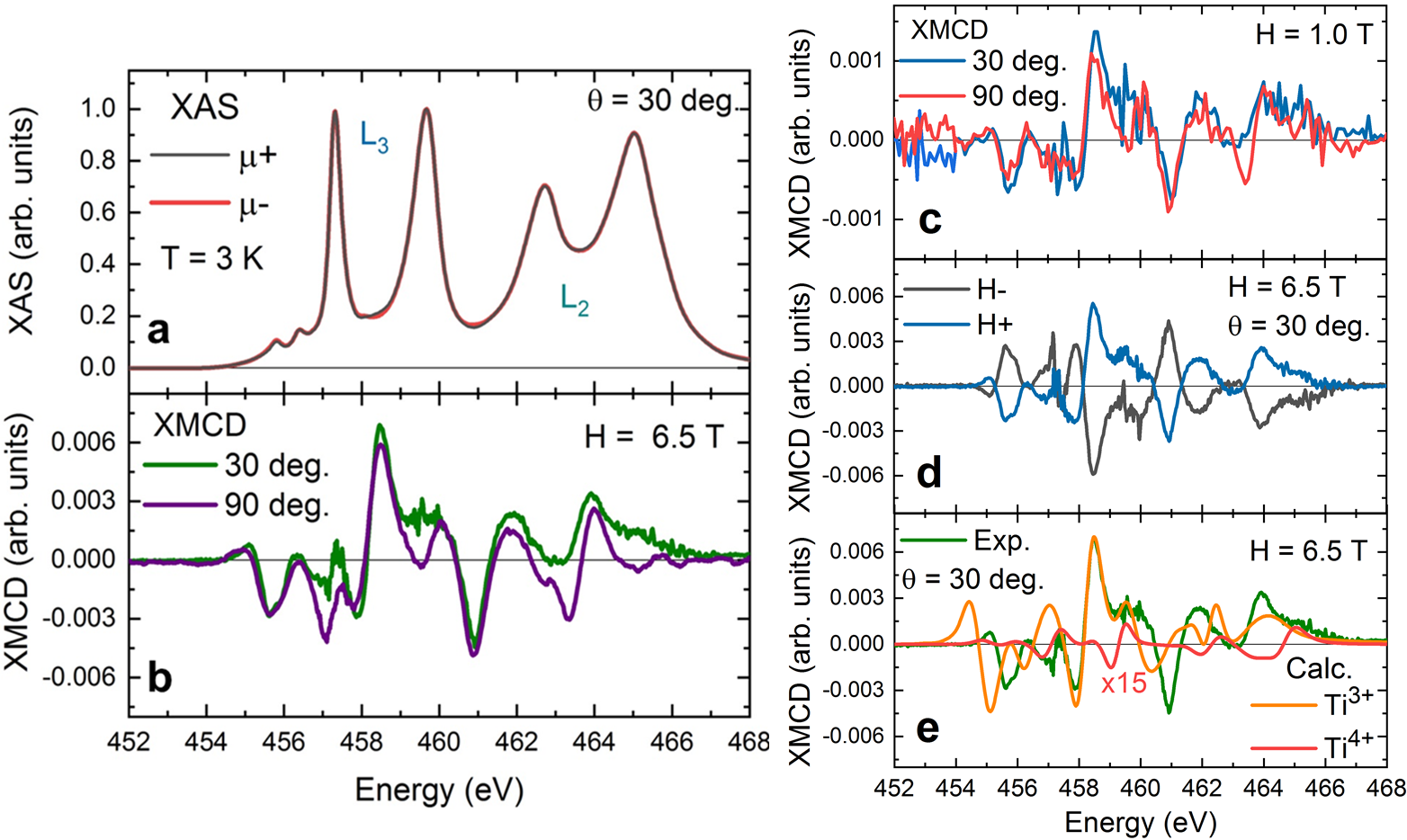}
\caption{(Color online) (a) XAS spectra acquired at 3 K with different photon helicities ($\mu_{+}$ and $\mu_{-}$) at $\theta = 30^{\circ}$. Panels (b) and (c) show the XMCD collected at $\theta = 30^{\circ}$ and $90^{\circ}$ under H = 6.5 T and 1.0 T, respectively. (d) XMCD signal for opposite magnetic field directions and (e) shows a comparison between experimental and calculated circular dichroic data based on extracted concentration of the two Ti oxidation states.  }
\label{fig:fig3_XMCD}
\end{figure}

Now we turn our attention to the magnetic properties of the system using XMCD.  
Figure~\ref{fig:fig3_XMCD}(a) shows XAS spectra taken with opposite photon helicities $\mu_{+}$ and $\mu_{-}$, in which the difference results in the XMCD signal (see panel~\ref{fig:fig3_XMCD}(b)) that has a maximum size of $\sim$0.6\% of the $L_3$ edge.   
The presence of a XMCD signal around the Ti $L_{2,3}$ edges gives clear evidence of magnetism at the interface.  
The observed reversal of the magnetic features for opposite field directions confirms the magnetic origin of the XMCD signal (Fig.~\ref{fig:fig3_XMCD}(d)). 
The panels~\ref{fig:fig3_XMCD}(b) and \ref{fig:fig3_XMCD}(c) also display the XMCD signal collected under 6.5 and 1.0 T, respectively. 
The similar magnetic contribution when the magnetic field is applied perpendicular ($\theta = 90^{\circ}$) or closer to parallel ($\theta = 30^{\circ}$) to the surface indicates a very small magnetic anisotropy of the system.\cite{XMCD_anisotropy} 
In addition, Fig.~\ref{fig:fig3_XMCD}(e) shows that the experimental XMCD spectra are in reasonable agreement with atomic multiplet calculations in which the signal is dominated by Ti$^{3+}$ magnetic moments.

\begin{figure}[!tb]
\centering

\includegraphics[trim=0.0cm 0.0cm 0.0cm -0.3cm, clip=true, totalheight=0.17 \textheight, angle=0]{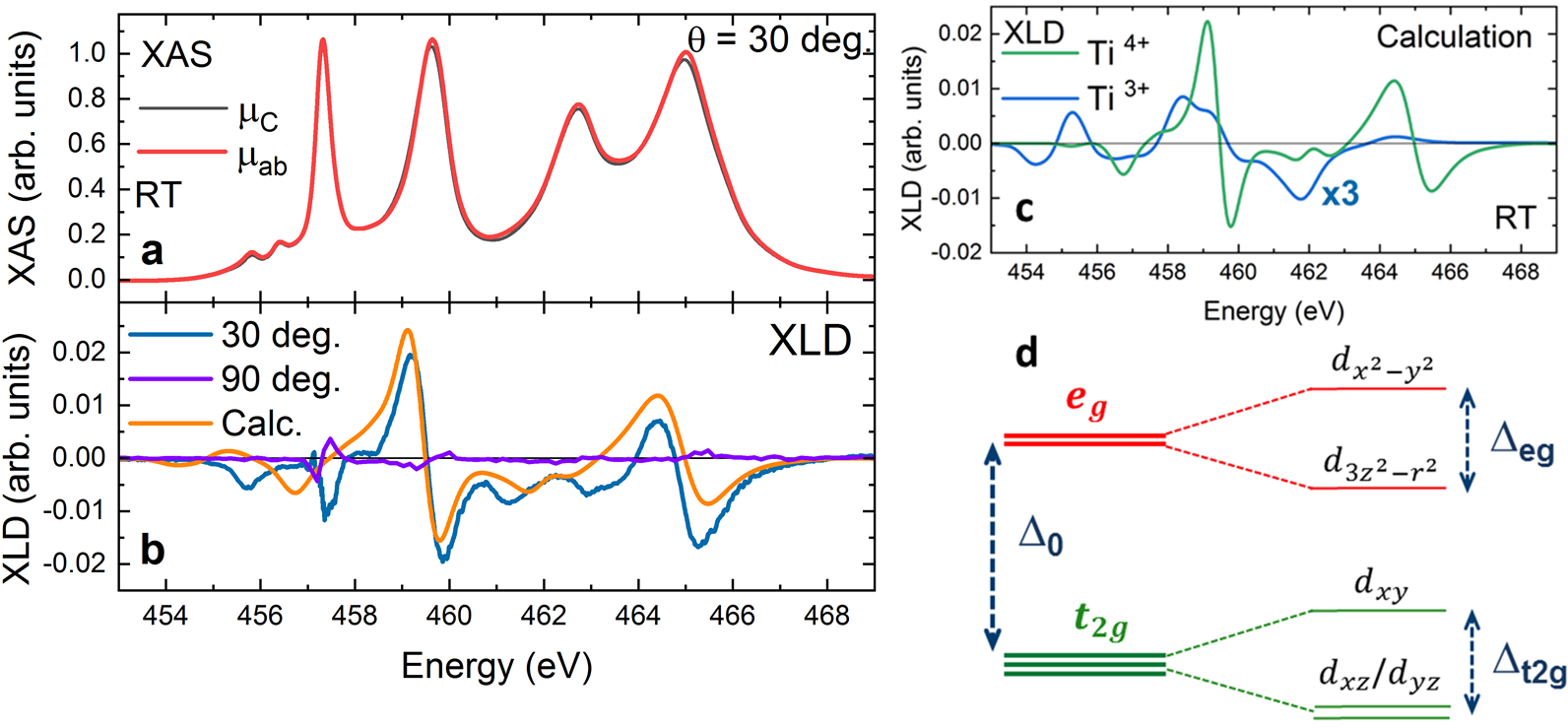}
\caption{(Color online) (a) XAS spectra obtained with horizontal and vertical  linear polarization probing $\mu_{c}$ and $\mu_{ab}$, respectively, at room temperature (RT) and at $\theta = 30^{\circ}$. (b) Calculated and experimental XLD intensities for grazing ($30^{\circ}$) and normal incident sample orientation ($90^{\circ}$). (c) XLD multiplet calculations assuming only Ti$^{4+}$ and Ti$^{3+}$ ions; a linear combination of the two spectra is applied in panel (b). (d) Schematics of the Ti$^{4+}$ 3$d^0$ orbital configurations at the GAO/STO interface.  }
\label{fig:fig4_XLD}
\end{figure}

The electron configuration of the Ti 3$d$ orbitals was investigated through the XLD data.  
Figure~\ref{fig:fig4_XLD}(a) displays XAS spectra with linear polarized light parallel ($\mu_{\text{c}}$) and perpendicular ($\mu_{\text{ab}}$) to the [001] direction collected at room temperature (RT).   
According to Fig.~\ref{fig:fig4_XLD}(b), the XLD signal is only present for $\theta = 30^{\circ}$ indicating an absence of in-plane anisotropy of the orbitals.    
The XLD signal collected in grazing incidence angle is much stronger than the XMCD signal (around 2$\%$ of the $L_3$ edge) and it is clearly observed at RT, indicating that it originates from the Ti crystal field splitting. 
This is because the XLD depends primarily on the crystal field splitting of the resonant atom which is little influenced by temperature.  
In addition, Figure~\ref{fig:fig4_XLD}(c) shows the calculated XLD spectra carried out separately for Ti$^{3+}$ and Ti$^{4+}$ ions. 
Assuming that the XLD is a linear combination of approximately 90  $\%$ of Ti$^{4+}$ and 10 $\%$ of Ti$^{3+}$ ions, good agreement is obtained with the observed XLD.  




Even though the features characteristics of the Ti$^{3+}$ ions are not distinctly observed in the XAS spectrum, the presence of Ti$^{3+}$ ions is confirmed by the atomic calculations and its good agreement to the XMCD/XLD data around the Ti $L_{2,3}$ absorption edges.  
The presence of these ions is crucial for the electronic and magnetic properties.  
As shown in panels~\ref{fig:fig3_XMCD}(b) and \ref{fig:fig3_XMCD}(c), the similar magnetic contributions in- and out-of-plane indicate a negligibly small magnetic anisotropy of the titanium magnetic moments.  
These results contrast the findings reported by Lee \textit{et al}. [\onlinecite{Lee_NatMat_2013}] for the LAO/STO interface, in which an absence of magnetic dichroism is observed when the magnetic field is applied perpendicular to the surface.    
However, the shape of the XMCD signal measured at grazing incidence angle agrees quite well with our data in which the magnetic signal originates from the Ti$^{3+}$ moments.
This raises several questions regarding a possible magnetic ordering and how these ions can influence the 2DEG at the interface.   

In order to investigate the origin of the magnetic signal, atomic multiplet calculations performed assuming a fourfold symmetry ($C_4$) with a Zeeman exchange field of 10 meV and charge transfer ($\sim$ 2 eV) are shown in Fig.~\ref{fig:fig3_XMCD}(e) for Ti$^{3+}$ and Ti$^{4+}$ ions. 
Magnetic features at the Ti $L_{2,3}$ edges are well described by the calculated XMCD spectrum considering the presence of localized Ti$^{3+}$ magnetic moments.  
 Salluzzo \textit{et al.} reported for LAO/STO interfaces that the origin of magnetism (the XMCD signal) can be achieved via two mechanisms:\cite{Salluzzo_PRL_2013,Mele_book_2015} due to the presence of localized Ti$^{3+}$ moments; or an asymmetry of the spin-up and spin-down electronic states in the 3$d$ band for Ti$^{4+}$ ions.
Multiplet calculations assuming the splitting of the 3$d^0$ (unoccupied) bands (see Fig.~\ref{fig:fig3_XMCD}(e)) suggest a signal more than one order of magnitude smaller than the Ti$^{3+}$ magnetic moments, which makes it difficult to observe it. 
However, a small amount of magnetic moments via splitting of the Ti$^{4+}$ 3$d$ bands can not be ruled out in our system.

\begin{figure}[!t]
\centering
\includegraphics[trim=0.0cm 0.0cm 0.0cm 0.0cm, clip=true, totalheight=0.18 \textheight, angle=0]{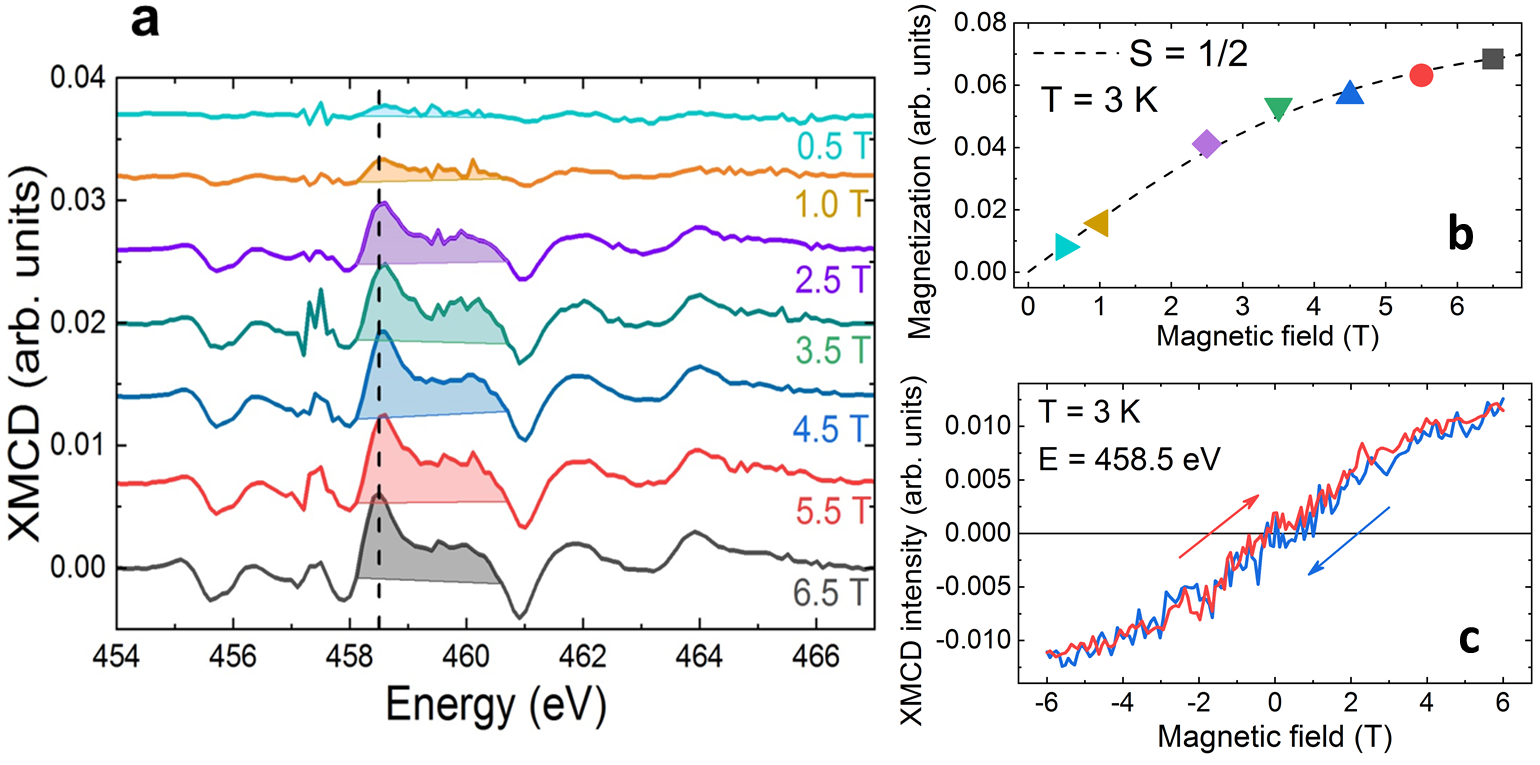}
\caption{(Color online) (a) XMCD signal as a function of external magnetic field. The spectra at different fields were shifted vertically for the sake of clarity. Panel (b) shows the evolution of the integrated intensity under the shaded area (between 458 and 461 eV) displayed in (a). Dashed line is a fit using (S = ${}^1/_2$) Boltzmann distribution. (c) Hysteresis loop obtained at the maximum of XMCD signal (dashed line around 458.5 eV in panel (a)). }
\label{fig:fig5_XMCD}
\end{figure}

After establishing the presence of localized Ti$^{3+}$ magnetic moments, we address the possible magnetic ordering at low temperature. 
Figure~\ref{fig:fig5_XMCD} shows the evolution of the XMCD signal as a function of magnetic field from 0.5 to 6.5 T.   
Since the ratio of the spin and orbital magnetic moment is expected to be temperature independent, the shaded areas between 458 and 461 eV (panel~\ref{fig:fig5_XMCD}(a)) in the XMCD signal provide a good estimate of the evolution of the magnetic moments.  
For lowering the magnetic field, the magnetization approaches zero, as can be seen in Fig.~\ref{fig:fig5_XMCD}(b), which confirms the absence of a remanent magnetization.  
The hysteresis loop shown in Fig.~\ref{fig:fig5_XMCD}(c) and performed at the maximum of the XMCD signal (dashed line in panel~\ref{fig:fig5_XMCD}(a): 458.5 eV) also indicates the absence of a remanent field.  
Therefore, any ferromagnetic ordering of the Ti$^{3+}$ magnetic moments must be approximately an order of magnitude smaller than the total free moments (either in size or fraction to the free moments) at the GAO/STO interface.   
In addition, the Fig.~\ref{fig:fig5_XMCD}(b) shows in dashed line the calculation for a quantum state of a half-integer paramagnet spin (S = ${}^1/_2$).     
Such a paramagnetic response as a function of field at this low temperature agrees quite well with our experimental results suggesting that the GAO/STO interface hosts localized paramagnetic Ti$^{3+}$ magnetic moments.  
As mainly discussed for LAO/STO system, beyond the trivial paramagnetic phase, its interface is proposed to develop a variety of magnetic order, such as ferromagnetism,\cite{Brinkman_NatMat_2007,Salluzzo_PRL_2013,Mele_book_2015} spiral-spin,\cite{Banerjee_NatMat_2013} electron pairing,\cite{Pai_arxiv_2016} etc.  
For GAO/STO and LAO/STO, Christensen \textit{et al.} [\onlinecite{Christensen_NatMat_2018}] observed striped shapes (dozens of $\mu$m in length) hosting long-range ferromagnetic order using scanning SQUID microscopy.  
However, these magnetic regions were not present homogeneously across the entire sample and they exhibited a strong temperature dependence where after thermal cycling these stripes were washed out or signals became weaker than the detection limit.  
Our x-ray absorption measurements were performed with an incident out-of-focus beam (spot size $\sim$ 1 x 0.5 mm$^2$), which would average over magnetic and non-magnetic areas. 
Hence, the overall paramagnetic response observed in XMCD not necessarily contradict the existence of magnetic ordered stripes at the GAO/STO interface.   

To elucidate the Ti 3$d$ electron configuration at the interface, XLD spectra were compared to atomic simulations.    
Crucial for the electronic properties is the position and the splitting of the $t_{2g}$ and $e_g$ states. 
Since the room temperature STO near the interface exhibits tetragonal symmetry, the 3$d$ subbands can behave differently and its ground state has major consequence for the electronic mobility of the 2DEG at the interface.      
Due to the four-fold symmetry normal to the [001] direction,\cite{Pesquera_PRL_2014, Cao_NPJ_2016} both linear polarized spectra are equal at normal incidence as seen in Fig.~\ref{fig:fig4_XLD}(b). 
Tiny features observed in the normal incidence XLD spectrum around 457-460 eV [panel~\ref{fig:fig4_XLD}(b)] are most likely artifacts due to the sharp peaks of the $t_{2g}$ and $e_g$ subbands.  
The good agreement between the experimental XLD data measured at $\theta = 30^{\circ}$ and the atomic simulation displayed in Fig.~\ref{fig:fig4_XLD}(b), allows an extraction of the crystal field splitting ($\Delta_0$), the splitting between the $d_{xy}$/$d_{xz, yz}$ ($\Delta _{t2g}$) and $d_{x^2-y^2}$/$d_{3z^2-r^2}$ ($\Delta _{eg}$) orbitals.  
The simulations were performed for both Ti valence states as displayed in panel~\ref{fig:fig4_XLD}(c), with best agreement found for dominating Ti$^{4+}$ signals. 
A crystal field splitting (see (panel~\ref{fig:fig4_XLD}(d)) of $\sim$1.9 eV with the $\Delta_{t2g}$ = 70 meV and $\Delta_{eg}$ = 310 meV  was found for the 3$d^0$ electron configuration at the GAO/STO interface.  
This indicates that the lowest energy levels are the out-of-plane $d_{xz}/d_{yz}$ subbands. 
These findings agree with Ref.~\onlinecite{Cao_NPJ_2016}, however, they are in contrast to the observations for the LAO/STO system in which the in-plane $d_{xy}$ subbands are lowest in energy. 
The lower energy of the $d_{xz}/d_{yz}$ subbands has direct impact on the orbital reconstruction at the interface, which highly affects the magnetotransport. \cite{Joshua_NatCommun_2012}
For the GAO/STO interface the $d_{xz}/d_{yz}$ subbands are at the lowest energy and there is an absence of magnetic anisotropy, $i.e.$, magnetic response with similar intensities when the magnetic field is applied in- or out-of-plane. 
This stays in contrast to the properties of the LAO/STO interface, where the $d_{xy}$ subbands being at lowest energy.  
In addition, the magnetic response is anisotropic.  
Although it was expected that the GAO/STO and LAO/STO heterostructures display similar properties, we demonstrated here that both present distinct electronic and magnetic properties.   
In addition, there is no straight forward connection between the anisotropy axis and the orbital level splitting that is defined by the crystal field, in particular for systems with very small angular magnetic moments as is the case here.  
Assuming though a fixed relation, the in-plane anisotropy observed in the LAO/STO interface occurring in $d_{xy}$ subbands, with the moment lying in plane of the orbital wave functions, would imply that for GAO/STO the moment will point equally along the $x$,$z$ ($y$,$z$) axis, respectively, being therefore isotropic as observed in the measurement.

In conclusion, our investigations using XAS techniques found a significant amount of the Ti$^{3+}$ states at the GAO/STO interface. 
The XLD measurements show that the ground state consists of $d_{xz}/d_{yz}$ subbands, that are only slightly lower in energy than the $d_{xy}$ subbands. 
XMCD shows that the Ti$^{3+}$ states are magnetic with moments being rather isotropic.
These moments show a dominant paramagnetic behavior at 3 K.  
The strong contrast to the LAO/STO interface suggests a fixed relation between orbital wave functions and magnetic anisotropy axis in these conductive interfaces.  \\



The X-ray absorption measurements were performed on the EPFL/PSI X-Treme beamline at the Swiss Light Source, Paul Scherrer Institut, Villigen, Switzerland. 
The authors are indebted to St. Zeugin, N. Daff\`{e}, and M. Studniarek for their assistance at the measurement setup, and A. Chikina, E. Bonini, M. Caputo, M. Radovi\`{c} and J. Dreiser for fruitful discussions. J.R.L.M. and S.P. were supported by the National Centers of Competence in Research in Molecular Ultrafast Science and Technology (NCCR MUST) from the Swiss National Sciences Foundations and N.O.H. and M.D. from the Swiss National Sciences Foundation Project No. $200021$-$169017$. D.V.C. and N.P. were supported by the NICE project, which has received funding from the Independent Research Fund Denmark, grant no. 6111-00145B.




\begin{thebibliography}{10}
\bibitem{geometry} XMCD (at H = 1.0 and 6.5 T) and XLD measurements were carried out in two different sample geometries: normal incidence ($\theta = 90^{\circ}$) or grazing incident angle ($\theta = 30^{\circ}$). The $\theta$ angle is illustrated in Fig.~\ref{fig:fig1_setup}(b) and (c). XMCD experiments performed at $30^{\circ}$ and $90^{\circ}$ provide information of magnetic moments in- and out-of-plane, respectively.
\bibitem{code_XAS} The simulations were performed assuming a $C_4$ symmetry with a crystal field parameters $\Delta_0=$1.9 eV, $\Delta_{eg}$ = 70 meV, $\Delta_{t2g}$ = 310 meV. The broadening of the Ti$^{4+}$ XAS spectrum was calculated using a Lorentz function with 0.03 eV ($L_3$, $t_{2g}$), 0.2 eV ($L_3$, $e_{g}$), 0.35 eV ($L_2$, $t_{2g}$), and 0.55 eV ($L_2$, $e_{g}$). For the XMCD calculations, a Zeeman exchange splitting of 10 meV was used as input.
\bibitem{XMCD_anisotropy} The small anisotropy between the XMCD signals measured at $\theta = 30^{\circ}$ and $90^{\circ}$ shown in Figure~\ref{fig:fig3_XMCD} (panels \ref{fig:fig3_XMCD}b and \ref{fig:fig3_XMCD}c) is most likely related to the normalization of the signal.
\end{thebibliography}
\end{document}